\input{aipcheck}

\documentclass[
    ,final            % use final for the camera ready runs
  ]
  {aipproc}

\layoutstyle{6x9}

\begin{document}

\title{ Pion Production in Neutrino-Nucleon Reactions}


\author{ E. Hern\'andez}{
  address={Grupo de F\'\i sica Nuclear,
Departamento de F\'\i sica Fundamental e IUFFyM,\\ 
Facultad de Ciencias, E-37008 Salamanca, Spain.}
}

\author{J.~Nieves}{
  address={Departamento de
 F{\'\i}sica At\'omica, Molecular y Nuclear,\\ Universidad de Granada,
 E-18071 Granada, Spain}}

\author{M.~Valverde}{
  address={Departamento de
 F{\'\i}sica At\'omica, Molecular y Nuclear,\\ Universidad de Granada,
 E-18071 Granada, Spain}}

\begin{abstract}
  We construct a model for the weak pion production off the nucleon,
  which in addition to the weak excitation of the $\Delta(1232)$ resonance
  and its subsequent decay into $N\pi$, it includes also some background
  terms required by chiral symmetry.  We re-fit the $C_5^A(q^2)$ form
  factor to the flux averaged $\nu_\mu p \to \mu^-p\pi^+$ ANL
  $q^2-$differential cross section data, finding a substantially
  smaller contribution of the $\Delta$ pole mechanism than traditionally
  assumed in the literature. We also show that the interference
  between the Delta pole and the background terms produces
  parity-violating contributions to the pion angular differential
  cross section. 
\end{abstract}

\maketitle

%%%%%%%%%%%%%%%%%%%%%%%%%%%%%%%%%%%%%%%%%%%%
%% MAINMATTER
%%%%%%%%%%%%%%%%%%%%%%%%%%%%%%%%%%%%%%%%%%%%

\section{Introduction}

The pion production processes from nucleons and nuclei at intermediate
energies are important tools to study the hadronic structure and play
an important role in the analysis of the present neutrino oscillation
experiments, where they constitute a major source of uncertainty in
the identification of electron and muon events. Therefore, it is
important to understand nuclear medium effects in the production of
leptons and pions induced by the atmospheric as well as accelerator
neutrinos used in these oscillation experiments. To this end, the
starting point should be a correct understanding of the reaction
mechanisms in the free space. In this talk,  we focus on
the weak pion production off the nucleon driven both by Charged and
Neutral Currents (CC and NC) at intermediate energies. The model
presented here will allow us to extend the results of
Refs.~\cite{ccjuan1} for CC and Ref.~\cite{ncjuan} for NC
driven neutrino-nucleus reactions in the quasielastic region, to
higher excitation energies above the pion production threshold up to
the $\Delta(1232)$ peak.

There have been several studies of the weak pion production off the
nucleon at intermediate energies~\cite{LS72}--\cite{Paschos}. Most of
them describe the pion production process at intermediate energies by
means of the weak excitation of the $\Delta(1232)$ resonance and its
subsequent decay into $N\pi$, and do not incorporate any background
terms.  In this talk, we also consider some background terms,
required by chiral symmetry. Their contribution\footnote{Some
  background terms were also considered in Refs.~\cite{FN}
  and~\cite{SUL03}. In the latter reference, the chiral counting was
  broken to account explicitly for $\rho$ and $\omega$ exchanges in
  the $t-$channel, while the first work is not consistent with the
  chiral counting either and it uses a rather small axial mass ($\approx 0.65$
  GeV), as well.} is sizeable even at the $\Delta(1232)-$resonance
peak and it turns out to be dominant near pion threshold. We re-adjust
the $C_5^A(q^2)$ form--factor that controls the largest term of the
$\Delta-$axial contribution, and find corrections of the order of 30\%
to the off diagonal Goldberger-Treiman relation (GTR)  when the ANL
bubble chamber cross section data~\cite{anl} are fitted. Such corrections
would be smaller if the BNL data~\cite{bnlviejo}
were considered. Additional results and all sort of
details can be found in Ref.~\cite{prd}.

\section{Theoretical Framework}
\begin{figure}
 \makebox[0pt]{\hspace{-0.3cm} \includegraphics[height=.15\textheight]
{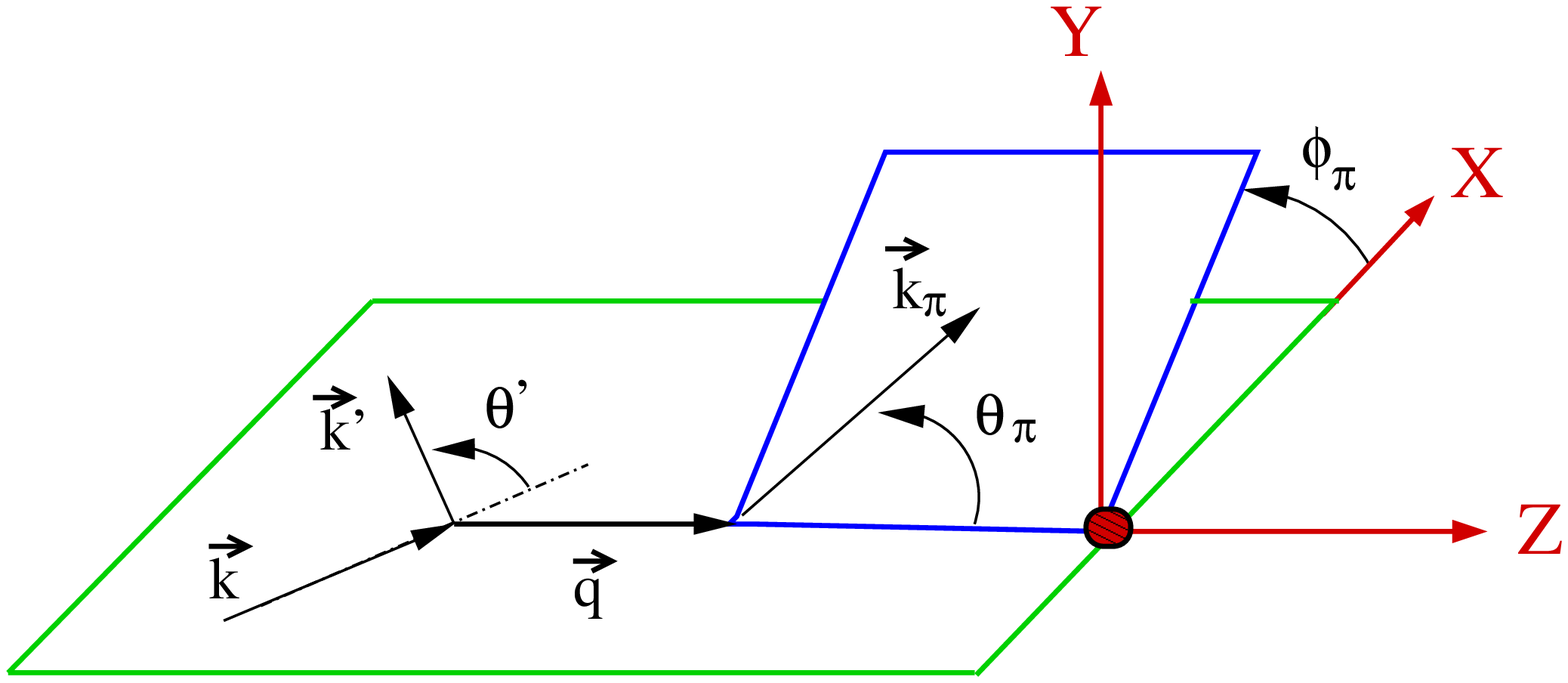}\hspace{0.25cm}
\includegraphics[height=.3\textheight]{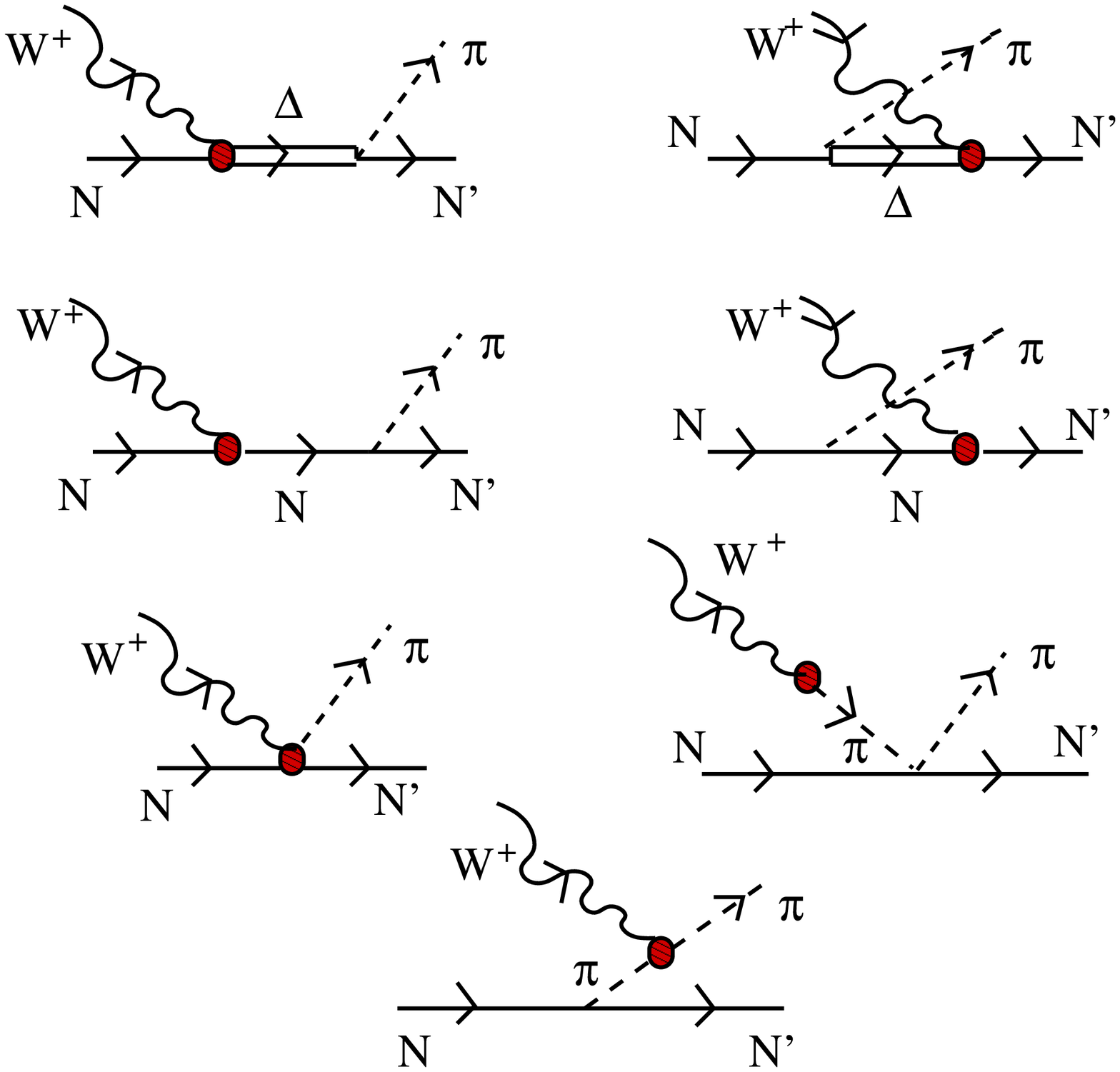}} 
  \caption{Left: Definition of the different kinematical variables
used through this work. Right:Model for the $W^+ N\to N^\prime\pi$
  reaction. It consists of seven diagrams: Direct and crossed
  $\Delta(1232)-$ (first row) and nucleon (second row) pole terms,
  contact and pion pole contribution (third row) and finally the
  pion-in-flight term.  The circle in the diagrams stands for the weak
  transition vertex.}
\label{fig:diagramas}
\end{figure}

We will focus on the neutrino--pion production reaction off the
nucleon driven by charged currents, $\nu_l (k) +\, N(p) \to l^-
(k^\prime) + N(p^\prime) +\, \pi(k_\pi)$, though the generalization of
the obtained expressions to antineutrino induced reactions and/or NC
driven processes is straightforward (see Ref.~\cite{prd}). The
unpolarized differential cross section, with respect to the outgoing
lepton and pion kinematical variables is given in the Laboratory
(LAB) frame (kinematics is sketched in Fig~\ref{fig:diagramas}) by
\begin{equation}
\frac{d^{\,5}\sigma_{\nu_l
    l}}{d\Omega(\hat{k^\prime})dE^\prime d\Omega(\hat{k}_\pi) } =
    \frac{|\vec{k}^\prime|}{|\vec{k}~|}\frac{G^2}{4\pi^2}
    \int_0^{+\infty}\frac{d|\vec{k}_\pi| |\vec{k}_\pi|^2}{E_\pi}
    L_{\mu\sigma}^{(\nu)}\left(W^{\mu\sigma}_{{\rm CC}
    \pi}\right)^{(\nu)} \label{eq:sec}
\end{equation}
with $G$ the Fermi constant and $L$ and $W$ the leptonic and hadronic
tensors, respectively. The leptonic tensor is given by
$\left[\epsilon_{0123}= +1, ~
g^{\mu\nu}=(+,-,-,-)\right]$: 
$L_{\mu\sigma}^{(\nu)} =
 k^\prime_\mu k_\sigma +k^\prime_\sigma k_\mu
- g_{\mu\sigma} k\cdot k^\prime + {\rm i}
\epsilon_{\mu\sigma\alpha\beta}k^{\prime\alpha}k^\beta $.
The hadronic tensor includes all sort of non-leptonic
vertices, 
\begin{equation}
  (W^{\mu\sigma}_{{\rm CC} \pi})^{(\nu)} =
  \overline{\sum_{\rm spins}}
  \int\frac{d^3p^\prime}{(2\pi)^3} \frac{ \delta^4(p^\prime+k_\pi-q-p)}{8ME^\prime_N}
  \langle N^\prime \pi |
  j^\mu_{\rm cc+}(0) | N \rangle \langle N^\prime \pi | j^\sigma_{\rm
    cc+}(0) | N \rangle^*
\label{eq:wmunu}
\end{equation}
with $q_\mu =k_\mu-k^\prime_\mu$, $M$ the nucleon mass and $j^\mu_{\rm
  cc+}$ the quark CC. Lorentz invariance
restricts the dependence of the cross section on $\phi_\pi$ (pion
azimuthal angle),
\begin{equation}
\frac{d^{\,5}\sigma_{\nu_l
    l}}{d\Omega(\hat{k^\prime})dE^\prime d\Omega(\hat{k}_\pi) } =
\frac{|\vec{k}^\prime|}{|\vec{k}~|}\frac{G^2}{4\pi^2}
     \left \{ A +
    B \cos\phi_\pi + C \cos 2\phi_\pi+
    D \sin\phi_\pi + E \sin 2\phi_\pi \right\} \label{eq:phipi}
\end{equation}
with $A,B,C,D$ and $E$  real, structure functions,
which depend on $q^2,\, p\cdot q$, $k_\pi\cdot q$ and $k_\pi\cdot p $. 
To compute the background contributions to the $\langle N^\prime \pi |
j^\mu_{\rm cc\pm}(0) | N \rangle$ matrix elements, we start from a
SU(2) non-linear $\sigma$ model involving pions and nucleons, which
implements the pattern of spontaneous chiral symmetry breaking of QCD.
We derive the corresponding Noether's vector and axial currents and
those determine, up to some form-factors\footnote{The form factor
  structure is greatly constrained by CVC and PCAC and the
  experimental data on the $q^2$ dependence of the $WNN$ vertex.}, the
contribution of the chiral non-resonant terms. To include the $\Delta$
resonance, we parametrize the $W^+ N \to\Delta$ hadron matrix element
as in~\cite{LS72} with the set of form-factors used in
~\cite{Paschos}. 
In particular, for the $C_5^A(q^2)$ form--factor that
controls the largest term of the $\Delta-$axial contribution, we use
\begin{equation}
C_5^A(q^2) = \frac{1.2}{(1-q^2/M^2_{A\Delta})^2}\times
\frac{1}{1-\frac{q^2}{3M_{A\Delta}^2}}, \quad {\rm with}\,~~ M_{A\Delta}=1.05
\,{\rm GeV}. \label{eq:ca5old}
\end{equation}
which leads to a reasonable description~\cite{Paschos} of the ANL
data. $C_5^A$ at $q^2=0$ is set to the prediction of
the off-diagonal GTR.  The  model consists of 7
Feynman diagrams (right panel of
Fig.~\ref{fig:diagramas}) constructed out of the $W N\to N$, $W N\to
\Delta$, $W N\to N\pi$,  $W \pi\to \pi$ and the $W\to \pi$ weak transition
vertices and the $\pi NN$, $\pi\pi NN$ and $\pi N \Delta$ strong couplings.
This model is an extension of that developed in Ref.~\cite{GNO97} for
the $ e N \to e' N \pi$ reaction.

\section{Results and Conclusions}
\begin{figure}[tbh]
\makebox[0pt]{\includegraphics[scale=0.59]{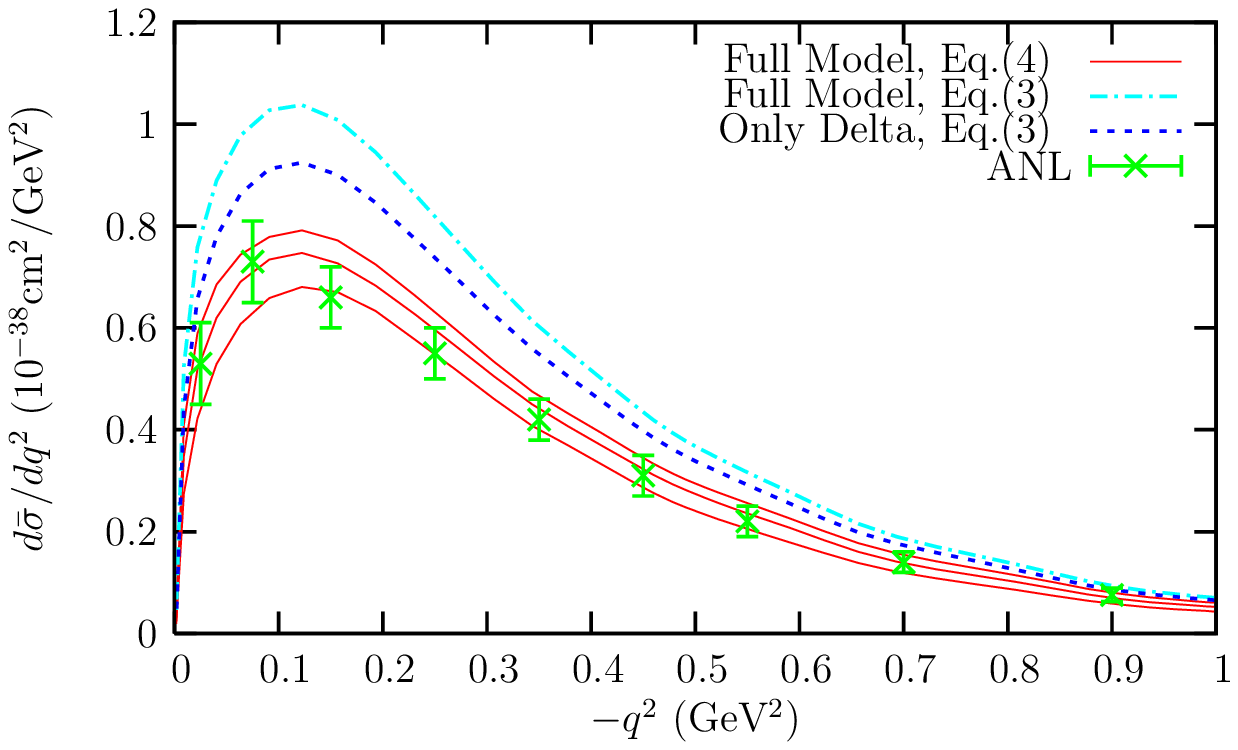}\hspace{0.2cm}\includegraphics[scale=0.59]{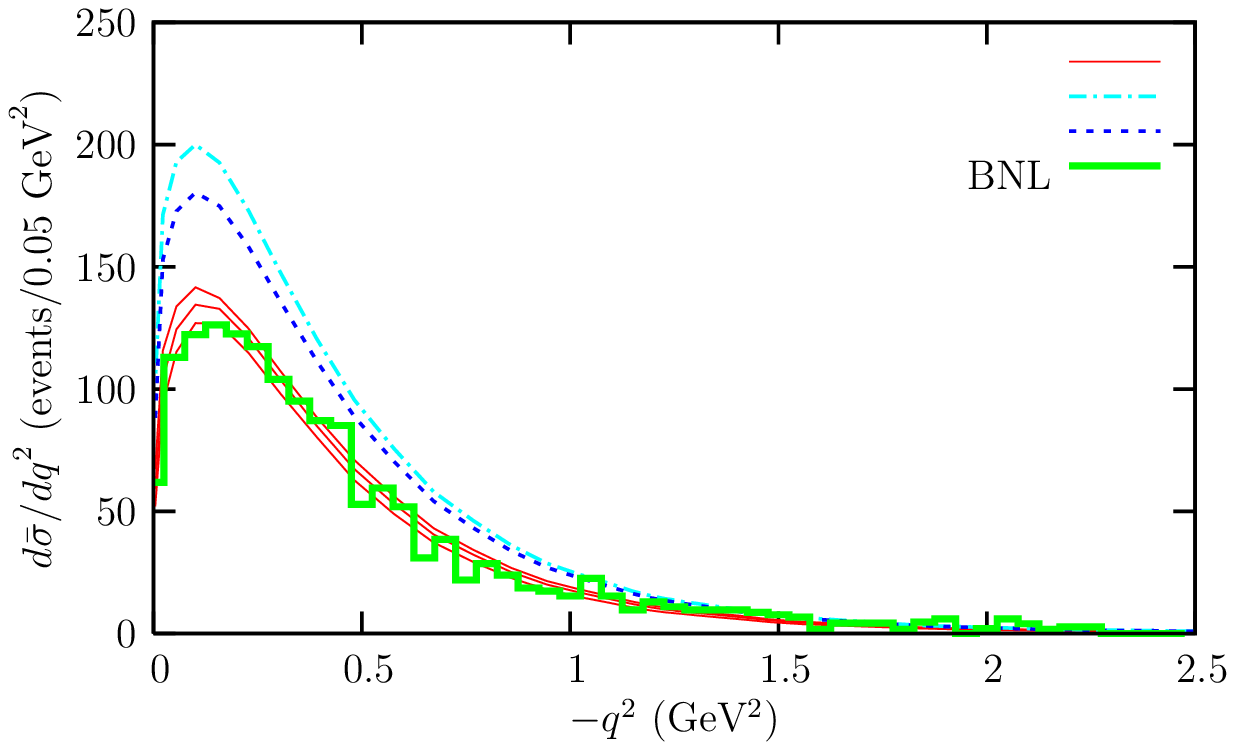}}
\caption{\footnotesize Flux averaged $q^2-$differential $\nu_\mu p \to
  \mu^- p \pi^+$ cross section $\protect\int_{M+m_\pi}^{1.4\,{\rm
  GeV}}dW \frac{d\,\overline{\sigma}_{\nu_\mu \mu^-}}{dq^2dW} $
  compared with the ANL~\cite{anl} (left) and BNL~\cite{bnlviejo} (right)
  experiments. Dashed lines stand for the contribution of the
  excitation of the $\Delta^{++}$ resonance and its subsequent decay
  ($\Delta P$ mechanism) with $C_5^A(0)=1.2$ and $M_{A\Delta}= 1.05$
  GeV. Dashed--dotted and central solid lines are obtained when the
  full model of Fig.~\ref{fig:diagramas} is considered with
  $C_5^A(0)=1.2,\, M_{A\Delta}= 1.05$ GeV (dashed-dotted) and with our
  best fit parameters $C_5^A(0)=0.867,\, M_{A\Delta}= 0.985$ GeV
  (solid). In addition, we also show the 68\% CL bands from 
Eq.~(\protect\ref{eq:besfit}).}\label{fig:anl-bnlq2}
\end{figure}

In Fig.~\ref{fig:anl-bnlq2} we present the flux averaged $q^2$
differential cross sections for the reaction $\nu_\mu p \to
\mu^-p\pi^+$ measured by the ANL and BNL experiments, with the $\pi N$
invariant mass cut $W\le 1.4$ GeV.  The agreement with the ANL data is
certainly worsened when the background terms, required by chiral
symmetry, are considered (dashed-dotted line). This strongly suggests
the re-adjustment of this form--factor. A $\chi^2-$fit  provides
\begin{equation}
C_5^A(0) = 0.867 \pm 0.075,~ M_{A\Delta}=0.985\pm 0.082\,{\rm
  GeV}, ~ ~\chi^2/dof=0.4~{\rm and}~r=-0.85 \label{eq:besfit}
\end{equation}
We observe a correction of the order of 30\% to the off diagonal GTR.
Results for total and differential neutrino cross sections, 
antineutrino and NC cross sections for different isospin channels can
be found in Ref.~\cite{prd}.  In general, the inclusion of the chiral
symmetry background terms brings in an overall improved description of
data as compared to the case where only the $\Delta P$ mechanism is
considered.  

The NC cross section dependence on $\phi_\pi$ constitutes a potential
tool to distinguish $\nu_\tau-$ from $\bar\nu_\tau$, below the
$\tau-$production threshold, but above the pion production
one~\cite{tau-nc}. The interference between the $\Delta P$ and the
background terms produces parity-violating contributions~\cite{prd} (structure
functions $D$ and $E$  in Eq.~(\ref{eq:phipi})) to
$d^{\,5}\sigma_{\nu_l l}/d\Omega(\hat{k^\prime})dE^\prime
d\Omega(\hat{k}_\pi)$, which might be used to constrain the axial
form factor $C_5^A$. 

\begin{theacknowledgments}
  This work was supported by DGI and FEDER funds, under contracts
  FIS2005-00810, FIS2006-03438, BFM2003-00856 and FPA2004-05616, by
  Junta de Andaluc\'\i a and Junta de Castilla y Le\'on under
  contracts FQM0225, SA104/04 and SA016A07.
\end{theacknowledgments}

\bibliographystyle{aipproc}   % if natbib is available

\end{document}